\documentclass[]{iscram} 

\usepackage[utf8]{inputenc}
\usepackage{lipsum}
\usepackage{tikz}
\usepackage{fancyvrb}
\usepackage{listings}
\usepackage{enumitem}
\usepackage{tcolorbox}
\usepackage{hyperref}
\usepackage{multicol}
\usepackage{siunitx}
\usepackage{xpatch}
\usepackage[super]{nth}
\usepackage{tabularx}
\usepackage{multirow}
\usepackage{nicefrac}       
\usepackage{microtype}      
\usepackage{xcolor}         
\usepackage{wrapfig}        
\usepackage{graphicx}       
\usepackage{subcaption}
\usepackage[japanese, russian, ukrainian, english]{babel}
\usepackage{CJKutf8}        

\lstdefinestyle{common}{
  xleftmargin=.5em,
  xrightmargin=.5em,
  frame=single,framesep=.5em,framerule=0pt,
  fancyvrb=true,
  basicstyle=\ttfamily,
  keywordstyle=\color{cyan!50!blue!75!black}\bfseries,
  commentstyle=\color{red!50!black}\itshape,
  stringstyle=\ttfamily\color{green!50!black},
  numbers=none,
  showspaces=false,
  showstringspaces=false,
  fontadjust=true,
  keepspaces=true,
  flexiblecolumns=true,
  emphstyle=\color{red},
}
\lstdefinestyle{TeX}{
  style=common,
  backgroundcolor=\color{blue!5},
  aboveskip=5pt,
  belowskip=5pt,
  language=[LaTeX]TeX,
  moretexcs={
    abstract, addbibresource, iscramset, keywords, mainmatter,
    maketitle, printbibliography, subsection, subsubsection, url,
    urldef, href, includegraphics, ldots, parencite, citeauthor,
    citeyear, citetitle, midrule, toprule, bottomrule
  },
  fancyvrb=true,
}
\lstdefinestyle{console}{
  style=common,
  backgroundcolor=\color{gray!10},
  aboveskip=5pt,
  belowskip=5pt,
}

\newlist{options}{description}{1}
\setlist[options]{%
  beginpenalty=10000,%
  itemsep=.5\parskip plus .3\parskip minus .2\parskip,
  parsep=.5\parskip plus .3\parskip minus .2\parskip,
  topsep=.5\parskip plus .3\parskip minus .2\parskip,
  partopsep=.5\parskip plus .3\parskip minus .2\parskip,
  style=nextline,labelindent=1em,%
  font=\normalfont\ttfamily}

\colorlet{macro color}{cyan!50!blue!75!black}
\colorlet{option color}{red!50!black}
\colorlet{generic color}{green!40!black}

\newtcolorbox{pseudoTeX}{colback=blue!5,colframe=blue!5,before=\nobreak}
\let\LaTeXorig\LaTeX
\renewcommand\LaTeX{\bgroup\fontfamily{lmr}\selectfont\upshape\LaTeXorig\egroup}

\urldef{\sitewebgind}\url{http://gind.mines-albi.fr}
\urldef{\sitewebminesalbi}\url{http://www.mines-albi.fr}
\urldef{\emailjohannes}\url{niujo64943@th-nuernberg.de}
\urldef{\emailmila}\url{mila.stillman@hm.edu}
\urldef{\emailanna}\url{anna.kruspe@hm.edu}
\urldef{\emailphilipp}\url{philipp.seeberger@th-nuernberg.de}

\urldef{\jdmail}\url{...@example.com}
\urldef{\sitex}\url{www.example.com}
\iscramset{
  CoRe Paper 2024={Social Media for Crisis Management},
  title={A dataset of Open Source Intelligence (OSINT) Tweets about the Russo-Ukrainian war
  },
  short title={OSINT Tweets about the Russo-Ukrainian war},
  author={
    short name={Niu},
    full name={Johannes Niu},
    affiliation={
      Technische Hochschule Nürnberg\\
       \href{mailto:niujo64943@th-nuernberg.de}{\emailjohannes}
    },
  },  
  author={
    short name={Stillman},
    full name=Mila Stillman,
    affiliation={
      Hochschule München\thanks{Research conducted while at Technische Hochschule Nürnberg}\\
       \href{mailto:mila.stillman@hm.edu}{\emailmila}
    },
  },  
   author={
    short name={Seeberger},
    full name= {Philipp Seeberger},
    affiliation={
      Technische Hochschule Nürnberg\\
       \href{mailto:philipp.seeberger@th-nuernberg.de}{\emailphilipp}
    }
  },
  author={
    short name={Kruspe},
    full name= {Anna Kruspe},
    affiliation={
      Hochschule München\\
       \href{mailto:anna.kruspe@hm.edu}{\emailanna}
    }
  }
}

\usepackage{tikzpagenodes}
\usetikzlibrary{fit}

\begin{document}

\maketitle


\abstract{Open Source Intelligence (OSINT) refers to intelligence efforts based on freely available data. It has become a frequent topic of conversation on social media, where private users or networks can share their findings. Such data is highly valuable in conflicts, both for gaining a new understanding of the situation as well as for tracking the spread of misinformation. In this paper, we present a method for collecting such data as well as a novel OSINT dataset for the Russo-Ukrainian war drawn from Twitter between January 2022 and July 2023. It is based on an initial search of users posting OSINT and a subsequent snowballing approach to detect more. The final dataset contains almost 2 million Tweets posted by 1040 users. We also provide some first analyses and experiments on the data, and make suggestions for its future usage.} 

\keywords{OSINT, Twitter, Intelligence, Social Media, Dataset} 

\section{Introduction}
Russia's ongoing war in Ukraine has sparked a surge of interest in Open Source Intelligence (OSINT) on online platforms such as Twitter. OSINT refers to the process of collecting, processing, analyzing and disseminating information that is publicly available, such as satellite images, live cams, leaked documents or journalistic reports. One of the main sources of OSINT in this war is the online content shared by soldiers, military bloggers or propaganda outlets from both conflict parties. Image and video material from the front lines often originates on platforms like TikTok or Telegram, where they are published by both military outlets and individual soldiers. On Twitter, a growing number of accounts collect, analyze, comment and share this material.

OSINT-related community efforts publish their findings under Twitter accounts such as @GeoConfirmed, or @oryxspioenkop. They base their analyses on material depicting the movement of troops or material that proves the destruction of military equipment. These analyses aim to generate insights into the nature and progression of current events along the lines of contact that go beyond the ``fog of war'' in a novel fashion. They also shape public narratives about the war: Prior to the full-scale incursion of Russian troops into Ukrainian territory on the February 24, 2022, OSINT-sourced posts highlighted the build-up of Russian military equipment and troops along the Ukrainian border during December 2021, January, and February 2022. This has likely contributed to the credibility of claims made by US intelligence agencies about a looming invasion \parencite{brantly2022narrative}. Similarly, both Russian and Ukrainian equipment losses tallied by the platform @oryxspioenkop ended up in media stories \parencite{pavlik2022russian}.

Social media plays a decisive role in the spreading of information, but also of propaganda and mis- or disinformation \parencite{euDisinfo}, and is particularly relevant in crisis situations \parencite{kruspe2021-actionable}. OSINT influences the way the war is portrayed, perceived and commented by an international audience. Like in other relevant public discourse spaces, propaganda and disinformation are likely used to influence public narratives \parencite{rode-hasinger-etal-2022-true}. Platforms such as Twitter, can become a potential source or amplifier of such disinformation spread via manipulated OSINT-related conversations.  
Understanding potential influences of propaganda or disinformation among OSINT-related accounts on Twitter requires insights into the content posted to Twitter by relevant accounts. This paper introduces a dataset of OSINT-related Twitter accounts and the top-level Tweets they published between January 1, 2022, and June 30, 2023, with the collection partially extended to July 2023. While not an exhaustive representation of the entire OSINT sphere on Twitter, the data enables comprehensive insights into OSINT related individuals and communities conversing about the Russian war against Ukraine and goes beyond other related data collections. A more detailed view of the dataset is presented in the Data analysis section, including the dataset's general and temporal statistics, language distribution and hashtag analysis, and a comparison to other datasets on the topic of the Russo-Ukrainian war. Additionally, we conducted a few experiments on the data, namely content analysis using a topic clustering approach, and misinformation detection, which are presented in the First experiments section.

\section{Related work}
In the context of the Russo-Ukrainian conflict, analysis of disinformation on social media has been a subject of research since the Euromaidan revolution of 2014 and the following open conflict in Crimea and Eastern Ukraine. \parencite{lange2015strategic} examines the strategic utilization of social media as a tool for information warfare in the context of the Russo-Ukrainian conflict. \parencite{RussDisinfoSocMed} explores disinformation campaigns conducted by the Russian state in general and a case study focusing on disinformation targeted at Ukraine during various stages, in particular during recent history. The Russian invasion in Ukraine on February 24, 2022, caused a surge of reports as well as image and video content posted to social media platforms. Due to the global scale and significance of the war, the analysis of this content has also increasingly gained attention. The role of OSINT in the Russo-Ukrainian war is explored in \parencite{kemp2023,flashpoint2023,tz2023}. \parencite{brantly2022narrative} examines the influence of OSINT-related content posted to Twitter on the narratives around the war. The survey focuses on Tweets published by a group of 24 hand-selected Twitter accounts well-known for publishing OSINT-related content and describes the potential impact of the reporting on Russian military movements before February 24, 2022.

In terms of data collection, an extensive attempt to preserve conversations about the war on Twitter was made by \parencite{kaggleDataset}. The dataset includes Tweets published between February 2022 and June 2023 extracted via predefined English hashtags and geolocation. \parencite{rode2023ukraine} presents a dataset of Tweets associated with war events described in other media. In \parencite{chen2023tweets}, a dataset of roughly 500M Tweets from February 2022 to January 2023 is presented. The authors queried the posts by keywords in English, Ukrainian, and Russian, and by using the Ukrainian flag emoji.
\parencite{park2022} collected data from Twitter and VKontakte posted by news outlet handles and media accounts between January 2021 and May 2022, including Russian ones. Other researchers created datasets as well, but focused their experiments on the area of content \parencite{ingole2023,Hanley_Kumar_Durumeric_2023a,Hanley_Kumar_Durumeric_2023b}, sentiment \parencite{poleksic2023, shevtsov2022twitter}, and misinformation analysis \parencite{pierri2023}. 

Most of these previous data collection approaches focus on hashtags, keywords, news outlets, or a small set of Twitter accounts.
In this work, we aim to create and analyze a comprehensive dataset targeted at OSINT-related accounts which enables further insights into public conversations about the war with a particular focus on OSINT communities. The ability to penetrate these communities to identify relevant users is key to enable such investigations. 

\section{Methodology}
The focus of this dataset are Tweets published by users that are connected to the term ``OSINT'' in the context of the Russo-Ukrainian war. We defined relevancy by applying the following two assumptions: 
\begin{enumerate}
    \item Relevant accounts are referenced (mentioned or retweeted) in conversations about OSINT in the context of the Russo-Ukrainian war, or mention other users in this context.
    \item Accounts publishing OSINT-related content converse with similar accounts via mentions or by retweeting their content.
\end{enumerate}
 
To collect Tweets from users of interest, we applied a two-step snowball sampling approach: In the first step we identified potentially relevant users by querying Twitter for Tweets that fit the first of the aforementioned criteria. The Tweets obtained in this step were not added to the final dataset, but served to identify relevant users. 
In a second step, we collected top-level Tweets from users deemed relevant in the first step. To increase the penetration rate of the population, we identified additional relevant users missed in the first step by applying the second criteria defined above to the Tweets collected from the first batch of relevant users. This approach is useful when suitable users could be hidden. 

\subsection{Identification of relevant users}
To identify relevant seed users, we conducted a keyword-based search, which queried both Tweets and replies that contained the term ``OSINT'' in combination with either the word "Ukraine" or "Russia". To discover multilingual results, both country names were translated to several European and Asian languages, which resulted in 15 and 23 further search terms for Ukraine and Russia, respectively. These languages include English, German, French, Spanish and other European languages as well as Korean, Japanese, Mandarin, and Hindu alongside Ukrainian and Russian. Multilingual search terms for the two countries were concatenated to a query $q = (c_1 \lor c_2 \lor \dots \lor c_n) \land \textrm{"OSINT"}$ with $c_i$ as the search term for one of the two countries.

This query was applied to a time frame ranging from July 1, 2021, until June 30, 2023, and resulted in 54,841 Tweets from 15,449 users. Since the Twitter search also yielded results if a search term was missing from the Tweet's text content but present in either the user handle (denoted by @) or user name (i.e. the search would yield a reply to a fictional user ``@osint\_user'', even if the text content of the reply only contains the term ``russia'', but lacks the required search term ``OSINT''), further filtering was required. This resulted in 32k Tweets posted by 9.5k users that contained any of the country search terms and the required term ``OSINT''. \autoref{tab:initial_search_metrics} depicts metrics of Tweets found in the initial search.

\begin{table}[h]
    \centering
    \caption{Tweet metrics of all Tweets found in the initial search.}
    \label{tab:initial_search_metrics}
    \begin{tabular}{lr}
        \toprule
        \textbf{Metric}      & \textbf{Count} \\
        \midrule
        Unique Tweets        & 32,477         \\
        Tweet Authors        & 9,581          \\
        Retweeted Users      & 1,558          \\
        Mentioned Users      & 2,187          \\
        \bottomrule
    \end{tabular}
    
\end{table}

\begin{figure}
    \centering
    \includegraphics[width=0.75\linewidth]{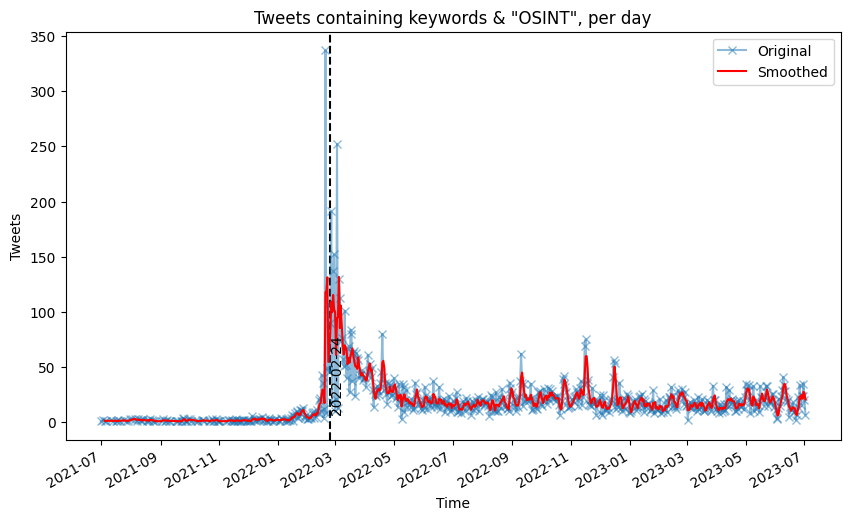}
    \caption{Number of Tweets retrieved in the initial search on a monthly basis.}
    \label{fig:initial_search_tweets_per_day}
\end{figure}

The temporal spread across the time frame can be seen in \autoref{fig:initial_search_tweets_per_day}. It  shows an uptick in interest in the topic starting from late January 2022, with an obvious peak around February 24, 2022, with a max of nearly 350 relevant Tweets per day. 

Following assumption 1, users of interest were defined as being referenced within these Tweets, either by being mentioned or retweeted, or by being the author of a Tweet about ``OSINT'' and either Ukraine or Russia, which applied to a total of 13,326 unique users in the set. To improve the results, we filtered these by applying a threshold: Only users who either authored or were referenced by at least ten Tweets that fit assumption 1 were considered relevant. This resulted in a list of 524 unique users. To limit this list to independent OSINT actors, these users were manually reviewed to remove accounts run by large media outlets (e.g. @bbc, @economist or @rt), as well as those of government officials or other institutions (e.g. @zelenskyyua, @potus, @unhcr or @ukraine). Due to the multilingual nature of the dataset, not all media outlets, government officials and institutions could be removed, as it would require a long research and matching of accounts. Moreover, the focus was mainly to remove large news outlets, therefore smaller news outlets could still be included in the dataset. This poses a limitation that could be studied and resolved in the future (e.g., by obtaining more information on official news Twitter accounts from different countries). In some cases, individuals may also mention or retweet Tweets from official news accounts. These were beyond the scope of this dataset. The remaining users were subsequently considered as relevant to this dataset and their Tweets collected, which constituted the second of the two-step approach as described in the Tweet collection section.

\begin{figure}
    \centering
    \includegraphics[width=0.7\linewidth]{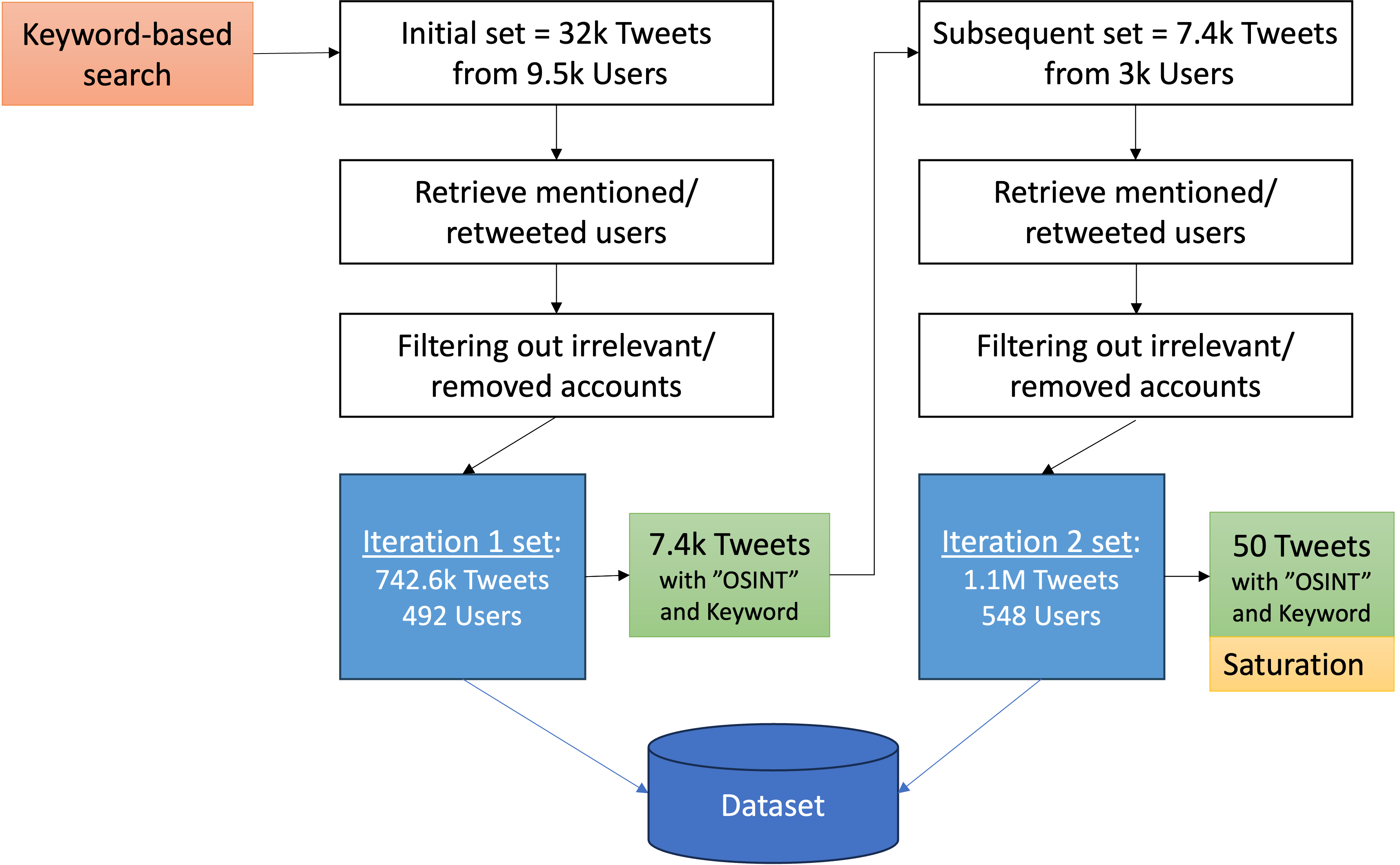}
    \caption{Snowball sampling approach with two iterations until reaching saturation for Tweets containing both ``OSINT'' and a country search term.}
    \label{fig:snowball_sampling}
\end{figure}

\subsection{Tweet collection}
We collected all top-level Tweets from users defined as relevant that were posted between January 1, 2022, and at least June 30, 2023. Due to technical reasons during the data aggregation and to ensure the completeness of data in the chosen time frame, the collection was extended into July 2023 for some users. To access historic Tweets, we used Twitter's search functionality. This led to two effects: First, since Tweets were collected with delay, interaction metrics are subject to bias, as older Tweets had more time to be interacted with than Tweets posted shortly before they were collected. Second, Tweets posted by users which had in the meantime been suspended, deleted, or had changed their profile settings to private could not be accessed. In other cases, the user handle used to identify users could not be resolved to a user. In the end, we collected top-level Tweets from 492 unique users. 
The collection yielded 742,583 unique Tweets published by these users, as \autoref{tab:combined_tweets_metrics} shows. 
200k retweets contained in the data reference 46k unique users. The Tweets further mention 44k users. 194,235 Tweets (26.16\%) contained one of the country-related keywords used for identification in the first step, indicating that a considerable part of conversations initiated by these accounts were indeed related to the Russo-Ukrainian war. However, only 7,434 (1\%) of Tweets also contained the term ``OSINT''. These Tweets were authored by or mentioned/retweeted 3,176 users. 
To identify additional users of interest, we applied the same assumptions to the collected Tweets that were used previously. We used the same criteria and filtering methodology as before, which reduced these 3k users to a list of 548 additional users. We subsequently collected top-level Tweets posted by these within the same time frame, as well, which yielded additional 1.1M Tweets. \autoref{tab:combined_tweets_metrics} shows the metrics of these.
While 366,318 (30.77\%) of the Tweets collected in this iteration contained at least one of the country search terms, only 50 also contained the search term ``OSINT'', indicating a saturation of the data. We therefore refrained from searching for additional relevant users to add to the dataset. The workflow using the snowball sampling approach is presented in \autoref{fig:snowball_sampling}.

The final dataset therefore contains 1,9M Tweets posted by 1,040 users. \autoref{tab:combined_tweets_metrics} shows that the data contains almost 500k retweets. The Tweets in this dataset also embed almost 600k images and 300k videos. 220k Tweets contain an external link.

\begin{table}[H]
    \centering
    \caption{Combined Tweet metrics of all Tweets in the dataset.}
    \label{tab:combined_tweets_metrics}
    \begin{tabular}{lll}
        \toprule
        \textbf{Metric}   & \textbf{Iteration 1} & \textbf{Iteration 2}  \\
        \midrule
        Unique Tweets   & 742,583  & 1,190,249  \\
        Tweet Authors   & 492      & 548        \\ 
        Retweeted Users & 46,916   & 43,700     \\ 
        Mentioned Users & 44,969   & 54,637     \\            
        Retweets        & 234,676 (31,6\%)  & 258,361 (21,7\%)    \\  
        Embedded Images & 205,322 (27,6\%)  & 392,846 (33,0\%)   \\ 
        Embedded Videos & 94,750 (12,8\%)   & 228,340 (19,2\%)   \\
        External Links  & 114,085 (15,4\%)  & 115,791 (9,7\%)   \\ 
        With keyword    & 194,235 (26,2\%) & 366,318 (30,8\%)   \\ 
        With Keyword and ``OSINT''  & 7,434 (1,0\%) & 50 (0,0\%)          \\
        \bottomrule
    \end{tabular}
\end{table}

To provide a first impression of the kind of users and Tweets contained in this dataset, we conducted a superficial qualitative survey of the data. Due to the scope and multilingual nature of the data, this provides only limited insights into the data. 
Accounts identified as relevant included those of journalists (e.g. @christogrozev, @ronzheimer), or military or political scientists (e.g. @ralee85, @timothydsnyder) that frequently tweet about the war. Other users appear to be connected to either the Ukrainian or Russian military (e.g. @tatarigamiua, @rybar\_en). Tweets in this dataset contain images or video footage, often sourced on Telegram or TikTok and distributed to Twitter, Other Tweets contain maps, satellite imagery or other reports. Tweets in this dataset contain commentary or analysis (e.g. geolocation) of events depicted in imagery. However, due to the focus on top-level Tweets, information or commentary contained in replies to Tweets, including threads (replies to one's own Tweets) are omitted. 
Other users featured in the dataset could be considered as general news aggregators, and would display a wider
variety of topics related to global or crisis related news. 
The dataset is publicly available under https://github.com/annakaa/OSINT\_Tweets.

\subsection{Significance and Limitations}

The use of the term ``OSINT'' allowed us to detect and penetrate the communities working on OSINT related to the Russo-Ukrainian war. By using this term, we were able to assure that the content of the Tweets is related to both OSINT and the Russo-Ukrainian conflict. However, there could be cases where similar topics were discussed and valuable information about the war was shared between users and the term ``OSINT'' was not explicitly used as a keyword or a hashtag. There could also be Tweets that use the term ``OSINT'' and discuss relevant topics without using one of the country search terms. The methodology of this paper is thus a rather restrictive one. It allowed us to capture some of these conversations, however it reached a saturation after the two iterations. 
Another bias was introduced during the snowball sampling method of collection, since it is based on mentions and retweets. This might limit the full spectrum of the discourse about the ongoing war within these communities. For example, replies and views could also contribute to the information spread. We also collected only users who were referenced (via mentions and retweets) at least ten times. This penalizing approach significantly reduced the number of relevant users. In the future, research could focus on changing and adjusting those parameters, adding softer intermediate filtering steps to collect an optimal number of relevant users and perhaps extending the number of iterations until reaching saturation. 

\section{Dataset Analysis} \label{datasetanalysis}
This section provides an overview of our first analyses of the dataset to characterize posts by the OSINT community, and also as the groundwork for future analyses. For a subjective impression of the data, the Appendix contains some examples of detected Tweets.

\subsection{Timeline, user reactions and attached media}

\begin{figure}
    \centering
    \includegraphics[width=0.7\linewidth]{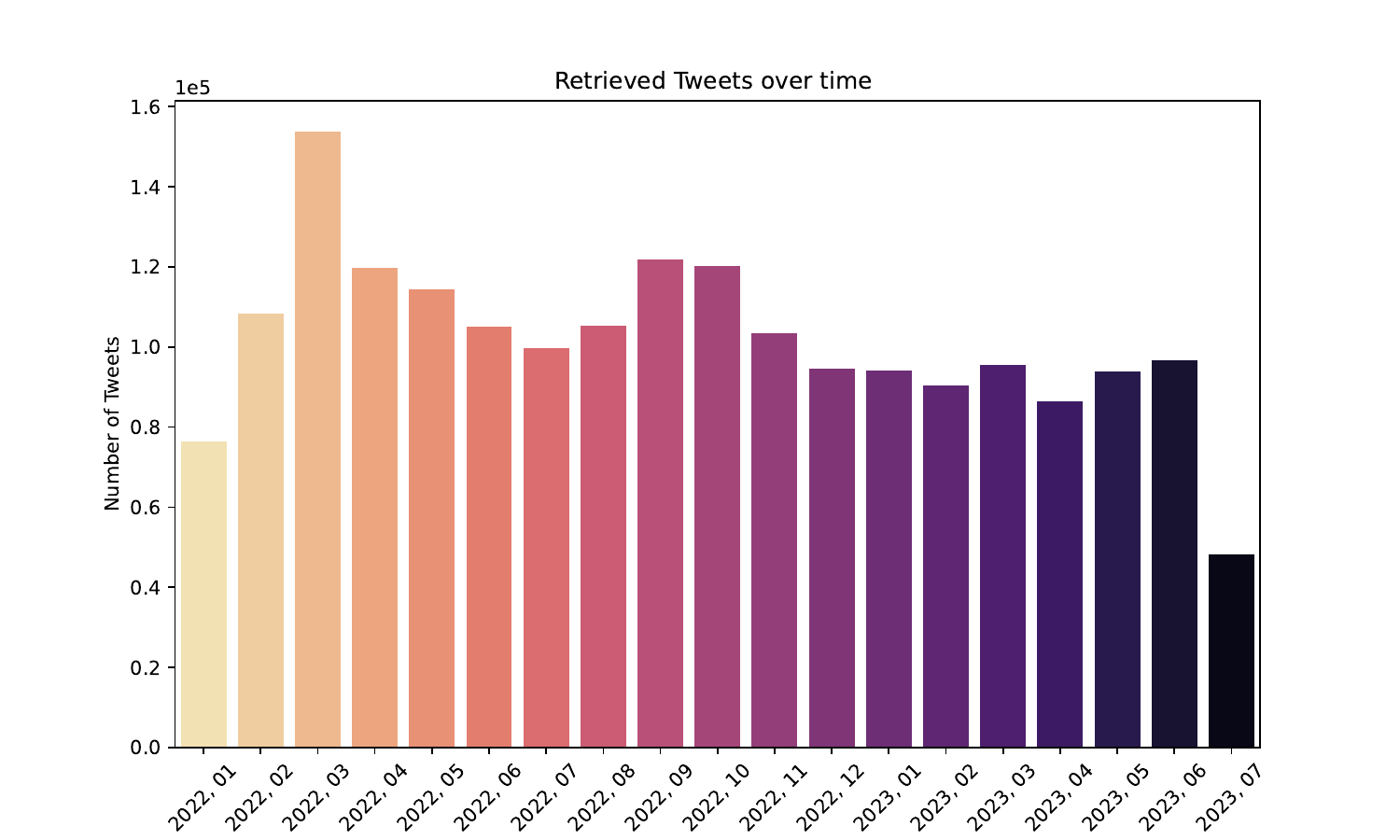}
    \caption{Number of collected Tweets in the dataset between January 2022 and July 2023 on a monthly basis.}
    \label{fig:tweets_per_day}
\end{figure}

\begin{figure}

 \begin{subfigure}[h]{0.45\textwidth}
     \centering
    \includegraphics[scale=0.45]{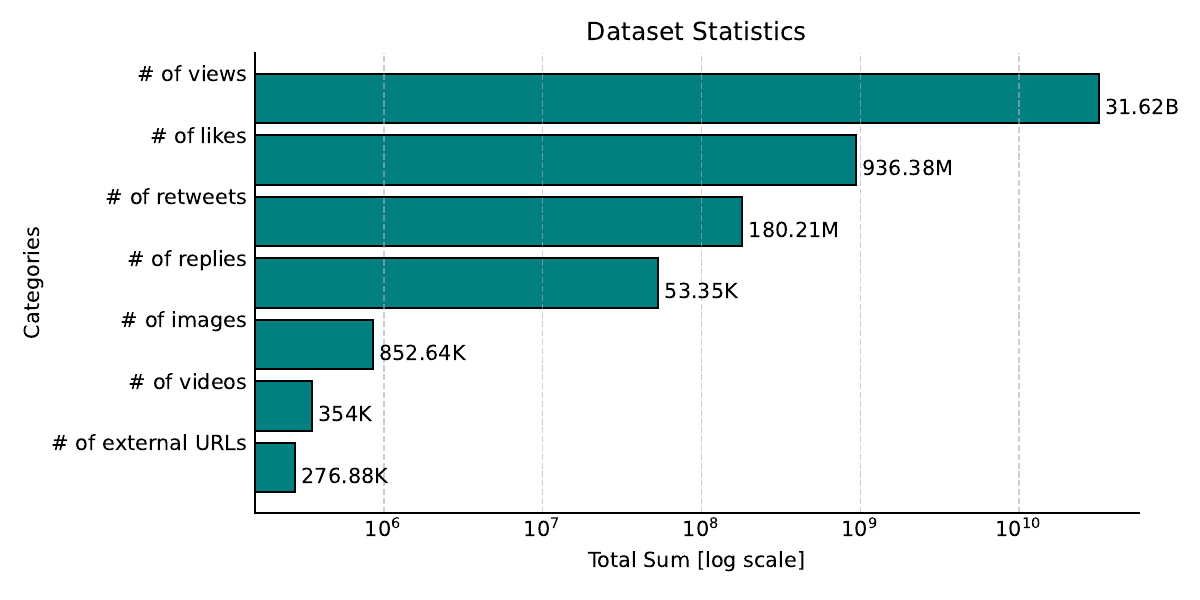}
\centering
\caption{General dataset statistics, including the total number of user reactions and attached media. Logarithmic scale was used for visualization purposes.}
\label{fig:stats}
     
 \end{subfigure}
 \hspace{2cm}
 \begin{subfigure}[h]{0.4\textwidth}
     \centering
     \includegraphics[scale=0.4]{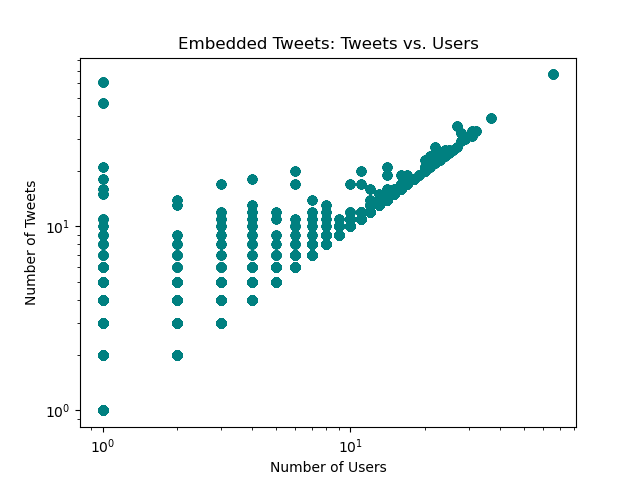}
     \caption{Embedded Tweets: number of unique user handles vs. number of Tweets containing the embedded Tweets. In logarithmic scale.}
     \label{fig:embedded}
     
 \end{subfigure}
 \caption{Dataset statistics}

\end{figure}

\autoref{fig:tweets_per_day} shows the monthly distribution of Tweets over time. As expected, the amount of Tweets within the OSINT community related to Russia and Ukraine peaked during the breakout of the conflict in March 2022, with around 160k Tweets in that month. In the following months, the number of Tweets declined, with another small peak in September and October 2022, plateauing at around 100k tweets per month until the end of the time frame. 
There are no major differences in the distribution of Tweets over time between the first and second iterations, therefore a combined timeline is presented. 
\autoref{fig:stats} depicts the total number of views, likes, retweets, replies, as well as additional media, i.e., images, videos and external URLs in the dataset. The user engagement in the community was quite high, with a total of 31.62B views, 936.56M likes, 180.3M retweets, and 53.3M replies. On average, that is about 16.3K views, 848 likes, 93 retweets and 27 replies per Tweet. Image links were included with around 600k Tweets, comprising 32.5\% of the dataset. Some Tweets contained multiple images which were added to the total image count. External URLs often indicated linked news articles or other social media posts (links to other Tweets were not considered here). The percentage of links to news articles was challenging to determine due to the multilingual nature of the dataset and, therefore, the large amount of reliable news sources that would have to be considered. Additionally, URLs of ``fake news'' web pages could contain the word ``news'' and affect the results. Determining specific channels of news and social media and their spread of information or misinformation could be studied in the future.

\subsection{Embedded Tweets}

About a quarter of the dataset contains embedded Tweets. 
Embedded Tweets, or retweets, allow another user to re-post a Tweet and comment on it. 
Embedded Tweets can be shared by any number of users with the same Embedded Tweet ID. Therefore, we counted the number of encapsulating Tweets and unique re-posting user handles for each unique Embedded Tweet ID. The results are shown in \autoref{fig:embedded}. 
This analysis revealed a general trend where Tweets were embedded in larger numbers often times by a few users (far left in the plot), often just one, typically within a short time frame. 
This can indicate that a user discussed another user's Tweet in detail. In some cases it was related to donation requests and rapid information spread during the outbreak of the war. In some cases, the text of the encapsulating Tweet was identical to the embedded one, with the embedded Tweet having a generic timestamp such as: '1900-01-01 00:00:00', suggesting potential bot activity. 
On the other hand, Tweets that were embedded by a larger variety of handles (moving to the right in the plot) indicated popular or controversial messages, with a large number of other users discussing, corroborating, or refuting them. Analyzing these embeddings further could be helpful for determining the legitimacy of users, spread of misinformation and potential scams.   

\subsection{Language distribution}

\begin{figure}[h]
\includegraphics[width=\textwidth]{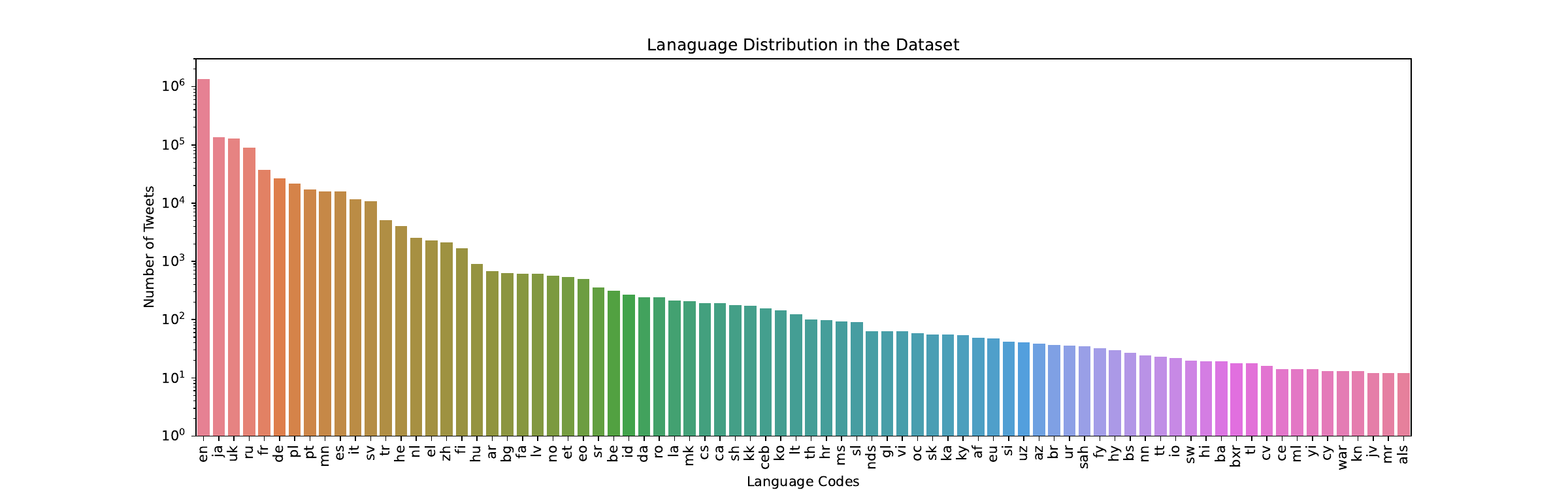}
\centering
\caption{Language distribution in the dataset according to the language codes of the \textit{fasttext} model. The 80 most frequent language codes are presented. Logarithmic scale was used for visualization purposes.}
\label{fig:language}
\end{figure}

We used the open-source library \textit{fasttext} \parencite{joulin2016}, which covers 176 languages in the ISO code format, to detect the language of each top-level and embedded Tweet with text content. Code-switching and Tweets with unidentifiable languages (e.g., very short ones, emojis or links only, etc.) were ignored. \autoref{fig:language} displays the language distribution of the top 80 languages in the final dataset. English (69\%) is the prevalent language, followed by Japanese (7\%), Ukrainian (6\%), Russian (5\%), German (2\%), and French (2\%). This aligns with our expectations and the distribution of other multilingual large-scale datasets \parencite{chen2023tweets}, apart from the large percentage of Japanese. Some of these user accounts represent Japanese news outlets which disseminate information about the Russo-Ukrainian conflict, but also many other topics. Additionally, the \textit{fasttext} model classified English Tweets in capital letters as Japanese Tweets, which resulted in about 6.5K Tweets mistakenly classified as Japanese. Furthermore, there were about 47K Tweets whose language was classified as 'unknown'. Overall, the dataset contains 147 languages. We did not find any significant differences for the language distribution of the embedded Tweets.

\subsection{Hashtags}
\begin{figure}
    \centering
    \includegraphics[width=\textwidth]{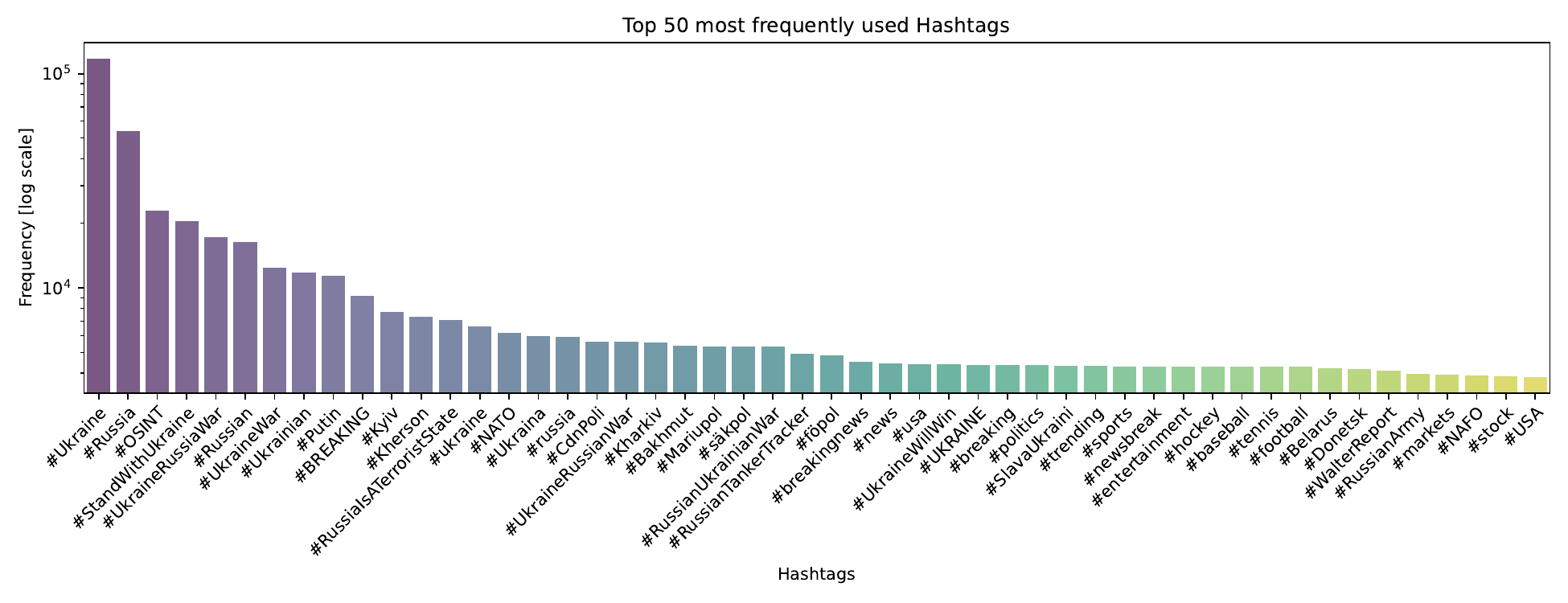}
    \caption{Top 50 hashtags used in the dataset and their frequencies. }
    \label{fig:all_tweets_top_hashtags}
\end{figure}

Hashtags are used frequently by users on Twitter to determine the topic of their post using short keywords. These can be used for data analysis allowing to label Tweets by topic. In addition, users often reuse hashtags, which allows to cluster the Tweets as well as to potentially make connections between users.
We see that these hashtags directly referred to topics related to the war or either country, including cities where battles occurred, and other news topics. The top two hashtags were \#Ukraine and \#Russia. The hashtag \#OSINT is the third most frequent one with around 25k occurrences. The hashtags (\#usa, \#NATO) can also be considered relevant to the topic. Interestingly, the top hashtags which were not related directly to the war often involved sports such as \#hockey, \#baseball, \#sports, etc. Observing the Tweets with sports related hashtags revealed that most of those Tweets belonged to a single account, which used all of those hashtags together when posting different types of news, both political and sports news. 
\autoref{fig:all_tweets_top_hashtags} shows the 50 most frequently used hashtags in the dataset. 

\begin{figure}[h]
\includegraphics[width=0.7\textwidth]{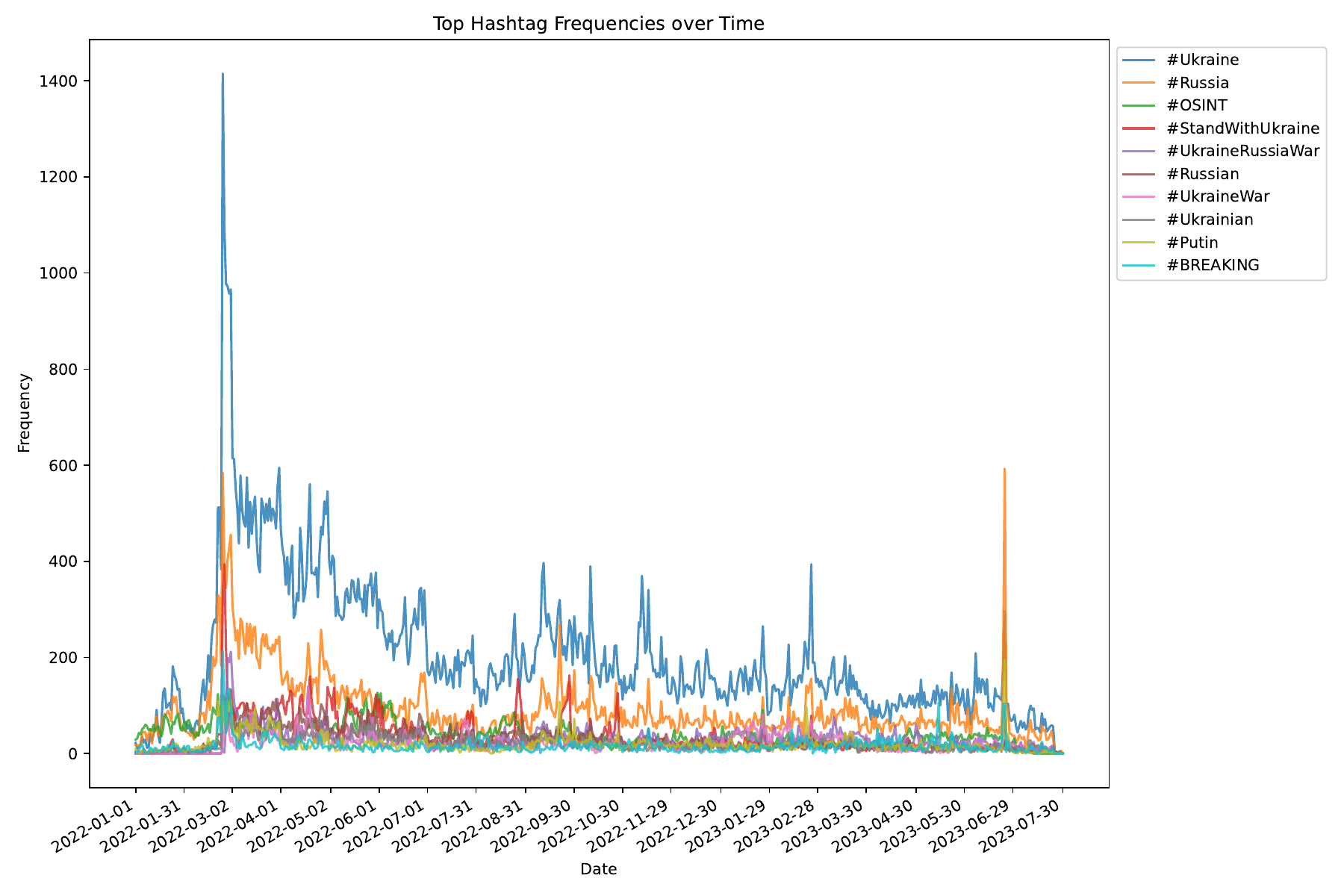}
\centering
\caption{Top 10 hashtags in the dataset over time on a daily basis.}
\label{fig:hashtags_over_time}
\end{figure}






\autoref{fig:hashtags_over_time} shows the temporal development of the top ten most frequently used hashtags on a daily basis. The overall trend of the two top hashtags \#Ukraine and \#Russia was similar to that of the whole dataset with the highest activity during the outbreak of the war and a gradual decline since then. Interestingly, their daily activity demonstrated similar fluctuations, suggesting that both hashtags were often used together. The hashtag \#Ukraine was used more frequently than \#Russia during the whole time period except for June 24, 2023, when the hashtags \#Russia, \#Putin and \#BREAKING spiked during the Wagner Group rebellion that began one day prior and was settled on that day. The hashtag \#StandWithUkraine had a few larger peaks between August and October 2022. The rest of the top ten hashtags had a similar pattern to the top two. 

\subsection{Comparison to other Datasets about the Russo-Ukrainian War}
Since some significantly larger datasets about the Russo-Ukrainian war have already been collected using other methods, we considered that those might capture the Tweets we have collected, and therefore, we compared a few of those with our dataset to assess its novelty and contribution. First, we compared the Tweet IDs of our dataset to the Tweet IDs of the Russo-Ukrainian War dataset \parencite{shevtsov2022twitter} available on GitHub. As of December 22, 2023, this dataset contained more than 123M Tweet IDs. We compared Tweet IDs due to their availability according to Twitter's policy of dataset sharing\footnote{\url{https://developer.twitter.com/en/developer-terms/policy}}. The analysis revealed no identical Tweet IDs existing in both datasets. 
Second, we compared our dataset to a large multilingual dataset available on Kaggle: the Ukraine Conflict Twitter Dataset  \parencite{kaggleDataset} using the Tweet and User IDs. This dataset comprises more than 70M Tweets and was collected using keywords related to the Russo-Ukrainian conflict, but nothing OSINT-specific.
Although our dataset is smaller in size compared to other datasets collected from Twitter on the same topic and around the same time-frame, the unique collection method that is based on OSINT and the engagement of users with the content, allowed us to discover 86 novel user accounts. Those accounts might be of interest, especially since they drive user engagement further, while not necessarily using specific keywords. Those hard-to-reach accounts might play an important role in the spread of both valuable information as well as misinformation. The total number of unique Tweet IDs in the dataset in this paper that differed from the unique Tweets IDs in the aforementioned dataset \parencite{kaggleDataset} was 1,764,340. Thus, only 168,492 Tweet IDs are identical in both datasets, making our dataset complementary. 
Additionally, an analysis of hashtags was performed for this dataset \footnote{\url{https://www.kaggle.com/code/josbenard/ukrain-war-hashtag-timeline?scriptVersionId=109668548}}. It turns out that the top twenty-five hashtags were not necessarily related to the Russo-Ukrainian conflict, including hashtags of other countries and world events. In our dataset, the top hashtags were mostly related to the war and to news in general, which demonstrates that although smaller in size, this method of collecting data created a focused and relevant dataset. This analysis suggests that both keyword-based data collection as well as through user interactions are valuable. A hybrid data collection approach could be explored in the future.

\section{First experiments} \label{firstex}
In this section, we present our first sample experiments on the data. These already provide interesting insights into OSINT communications about the war, but also serve as a basis for future research using our dataset.

\subsection{Content analysis}
In order to provide a first comprehensive overview of the topics covered in our dataset, we analyzed the textual content of the Tweets. 
However, we still expect irrelevant content related to specific accounts such as private users, journalists, political analysts or general news aggregators. 
These accounts provide content beyond the information content w.r.t. war, including other geopolitical events, personal opinions, reactions, political speeches, and advertisements. 
Since we want to focus on topics directly related to the Russo-Ukrainian war, we approach the content analysis with two consecutive steps. 
The first step implements the necessary filtering by conducting relevance detection as a binary classification task. 
The positive class corresponds to posts that are relevant to the Russo-Ukrainian war, while the negative class corresponds to the remaining posts. 
In a second step, we perform the clustering of relevant posts in order to investigate the topics within the dataset.

For relevance detection, we randomly sampled 1,000 English posts from the entire dataset and manually categorized the samples into “Relevant” as the positive and “Irrelevant” as the negative class. 
However, the classification approach and relevance of a post is determined by the information need of a potential user that differs depending on the use case and research objectives \parencite{kruspe2021-actionable,seeberger2022}. 
In this work, we annotated posts as relevant if they contain information about incidents and events directly affecting the war. 
This includes but is not limited to weapon deliveries, combat operations, troop and equipment movements, casualties and losses, and similar events in the Ukraine, Russia as well as Ukraine’s neighboring countries.
With the aim of enhancing situational awareness, which assists in understanding the 'big-picture' during crisis situations for end-users such as decision makers and emergency responders \parencite{vieweg2010microblogging,Olteanu2014CrisisLexAL}, we consider posts irrelevant if they are not related to the Russo-Ukrainian war or only contain personal opinions, sentiments, reactions, and political speeches. Those were shown to appear often in on-topic Twitter posts, but have not necessarily contributed to situational awareness.
This annotation process resulted in 346 relevant and 654 irrelevant posts. 
For the detection task, we trained a BERT-based\footnote{\url{https://huggingface.co/roberta-base}} classifier \parencite{devlin2019} and evaluated the model with stratified 5-fold-cross-validation (0.815 Precision, 0.899 Recall, and 0.854 F1-score).
We also experimented with larger BERT-based models and task-specialized sentence transformers \parencite{seeberger2023} but did not obtain substantial improvements.

For topic clustering, we filtered the dataset with the trained classifier by predicting the class for each English post, resulting in 287k (20.31 \%) relevant and 1126k (79.96 \%) irrelevant samples. 
We utilized Uniform Manifold Approximation and Projection (UMAP) \parencite{mcinnes2018} for the dimensionality reduction of Sentence-BERT embeddings \parencite{reimers2019} and clustered the relevant posts with Hierarchical Density-Based Spatial Clustering (HDBSCAN) \parencite{mcinnes2017}. 
Only clusters with at least 200 elements were considered, resulting in a total of 142 clusters. 
As the number of clusters remained relatively high, we selected five clusters from the hierarchy to show the diverse topics covered in our dataset. 
In \autoref{tab:content_analysis}, we present the five clusters with high-level topics and representative Tweets. 
The identified topics range from coarse-grained topics such as the tracking of destroyed equipment to more fine-grained topics such as the Kakhovka dam collapse as local event. 

However, it is important to note that we found prediction errors in both the relevant and irrelevant subsets.
These prediction errors may be influenced by the small amount of training data for the wide range of topics.
To identify erroneous categories, we additionally conducted the clustering experiments on the irrelevant subset and highlight potentially false positive and false negative predictions. 
The majority of false positives concern geopolitical reports about other conflicts such as in Somalia, Afghanistan, Israel, and Philippines or events such as spy ballons, balistic missile tests, and shootings. False negatives frequently include specific locations, weapon systems, and posts that redirect to media footage without providing sufficient information in the textual content.
We refrain from analyzing the error cases in detail at this point and leave this to future work, as this is closely linked to the respective annotation process and the research objectives.

\begin{table}[ht]
    \centering
    \caption{Content analysis of five selected clusters with relevant posts. The topics describe the high-level information content of the cluster. For each cluster, we selected one representative tweet to highlight the information content.}
    \label{tab:content_analysis}
    \begin{tabular}{p{0.3\linewidth}p{0.65\linewidth}}
      \\\toprule
      \textbf{Topic} &  \textbf{Representative Tweet}
      \\\midrule
      Destroyed and captured equipment & \#UkraineWar: Newly added Russian equipment losses: 1x BMP-3 IFV (destroyed) 1x BTR-82A IFV (destroyed) 1x BTR-80 APC (captured) 1x Pontoon bridge (destroyed) 1x KamAZ truck (destroyed) 1x Ural-4320 truck (destroyed)
      \\
      \\
      Casualities and losses & Russian losses, 24 Feb 2022 to 8 Apr 2023: ~177,680 killed 3,636 tanks 7,020 armoured vehicles 2,727 artillery systems 533 MLRS 282 air defence systems 307 aircraft 292 helicopters 2,298 UAVs 911 cruise missiles 18 ships/boats 5,599 vehicles/fuel trucks 304 special equipment
      \\
      \\
      Airplane and ship tracking & \#USAF RQ-4A Global Hawk very close to \#Crimea on ISR mission, as a \#RuAF TU-154B is leaving \#Sevastopol. This maybe means an imminent amphibious assault by \#Russia. \#Ukraine \#OSINT
      \\
      \\
      Geolocalisation & "Probably an ammonium nitrate tank exploded in Dovhen'ke" Not Chlorine gas. 49.021734, 37.308826 GeoConfirmed by @Wolltigerhueter \#GeoConfirmed
      \\
      \\
      Kakhovka dam collapse & The water level at the Nova Kakhovka dam is up by 5 metres and several islands downstream have already been flooded, Russian-backed mayor of Nova Kakhovka says - TASS
      \\\bottomrule
    \end{tabular}
    
\end{table}


\subsection{Misinformation detection}
In a first attempt to identify fake news and misinformation in the collected Tweets, we conducted two simple experiments by closely examining the textual contents of all English language Tweets (1.1M) in the dataset. We used two different pre-trained text sequence classifier to identify Tweets that might contain false information.

In a first experiment, we tested a pre-trained model for COVID-19 misinformation detection\footnote{\url{https://huggingface.co/spencer-gable-cook/COVID-19_Misinformation_Detector}} and obtained 207,227 (17.5\%) Tweets that were classified as ``misinformation''. In a second experiment, we used a model trained on a more general ``fake news'' dataset\footnote{\url{https://huggingface.co/FriedGil/distillBERT-misinformation-classifier}}, which labeled 14,244 (1.06\%) Tweets as ``unreliable'' based on the text sequence. 

The results of both approaches are subject to biases: In both cases, we have no way to determine the accuracy of the results. The model used in the first experiment trained distillBERT to detect COVID-19 related misinformation in text sequences, which probably has a negative impact on its ability to generalize. A first manual review of Tweets labeled as unreliable in this experiment suggests, that positives were distributed across the dataset proportionally to the general volume of Tweets; we analyzed the temporal spread as well as the user and hashtag dissemination in positive Tweets and found that they resemble the general dataset. 

The second experiment yielded more insightful results. The model fine-tuned distillBERT to conduct based text misinformation classification. In this case, however, we also have no way of determining the accuracy of these findings without further work. 

\section{Conclusion and Future Work}
The dataset introduced in this work represents a portion of the ongoing public conversation about the Russo-Ukrainian war on Twitter. While by no means exhaustive, it enables targeted insights into the OSINT community on Twitter.

It is part of the nature of warfare that nobody has a full picture of the current and ongoing situation because the affected parties may not want the general public to know about their actions for various reasons, but also because the complex network of events simply makes it impossible to track all of them at all times. This problem is known as the ``fog of war'' \parencite{bjola2022information}. The issue is further complicated by the intentional or unintentional spread of misinformation. We are therefore reliant on various information sources inside the situation as well as outside, both while those events are unfolding and later on. During a crisis situation, independent actors and individuals share content about the event, whether as a way of showing support, asking for donations or spreading misinformation. Those independent players often engage with each other, driving the discourse further and reaching a larger audience. According to our findings, the snowball sampling approach was beneficial to the discovery of novel related user accounts and Tweets, extending beyond a simple keyword search. The snowball sampling could be improved in the future, perhaps by allowing less penalizing criteria to increase the amount of collected data and the number of iterations until reaching saturation. We hope that this dataset and our first experiments could provide a contribution to this end. 

It is, of course, important to also consider ethical issues in this context, such as data and user privacy \parencite{zhu2022} or even the safety of users distributing novel (and perhaps controversial) information. According to \parencite{rossi2022unwinding}, due the GDPR, research performed on social media data within Europe is considered similar to research with human subjects. It is therefore important to process and store this data carefully and to provide anonymity for social media users, to avoid risking the exposure of individuals without their given consent. This is especially important for accounts that have been deleted or suspended after the time of the data collection. Such methods have been explored in \parencite{rossi2022unwinding}, including encrypting stored data and anonymization techniques. This includes avoiding publishing user handles with explicit texts of Tweets and ideally using only aggregated statistics. The latter would be the only way to ensure that no individuals could be identified, and thus making the dataset fully anonymized. In this paper, we attempted to adhere to the ethical recommendations, and only publish a small amount of user handles of accounts with a large following, that could be considered public figures. We also demonstrated texts of representative Tweets as  examples of a few results of the content analysis experiments, without mentioning the users who posted them. However, this is a topic that would require further research, as well as a consultation with an official ethics department or institution, since there is not a simple established protocol for researchers to follow on ethical social media data mining. 

Possible future applications of the dataset include a combination and comparison with other data sources, such as traditional media or remote sensing data where it is not already part of the OSINT analysis. It also enables modeling of information spread on a temporal basis as well as a geographic one (i.e. what information is revealed where and at what point in time). Looking at the data itself, differentiation between ``correct'' OSINT and misinformation distributed under the same label would be of high interest \parencite{rode-hasinger-etal-2022-true}. This could be connected to user network analyses within this domain. Novelty detection could be applied to determine where information was first published \parencite{kruspe2021-novelty}. Additionally, OSINT data can be used to determine social opinion and emotions, crime and propaganda detection, detection of deepfakes and improving protection against cyberattacks \parencite{pastor2020not}. 
Beyond the dataset itself, the developed methodologies for its construction and exploration could be applied to other topics that spread dynamically on social media and pose similar problems for analysis, including other crises, but also social phenomena like \textit{\#metoo} or the antivax movement. The study of such movements on social media using an adapted snowball sampling approach could assist in gaining a deeper understanding of information spread, and the stages at which such movements could reach large audiences and become influential.


\newpage
\printbibliography

\newpage
\appendix




\section{Example Tweets}\label{app:examples}
	
\begin{table}[ht]
    \centering
    \begin{tabular}{p{0.95\linewidth}}
    \hline
    \\
        \textbf{https://twitter.com/TreasChest/status/1605953315065327617} \\\\
        As of now, the "Engels-2" military airfield is concentrated: — 20 strategic missile-carrying bombers Tu-95ms (Bearnet) — 6 Tu-160 (Blackjack) strategic missile-carrying bombers;) — 1x Il-76 (Candid) heavy military transport aircraft \\\\
    \hline
    \\
        \textbf{https://twitter.com/Flash\_news\_ua/status/1659216874901770240} \\\\
        During the day, the enemy launched 36 missile and 23 air strikes, as well as about 30 attacks from MLRS on the positions of our troops and populated areas, — the General Staff of the Armed Forces of Ukraine.\\\\
    \hline
    \\
        \textbf{https://twitter.com/front\_ukrainian/status/1604507300403613703} \\\\
        Ukrainian long range MLRS Himars have hit 8 personnel concentration areas, 2 control points, and 4 ammunition depots of Russian forces over the past day.\\\\
    \hline
    \\
        \textbf{https://twitter.com/Hromadske/status/1518869992908935168} \\\\
        From the beginning of the full-scale invasion of Ukraine until this morning, Russia lost about 22,100 military personnel, 184 planes, 154 helicopters, 8 naval vessels, 2.308 armored combat vehicles, and 919 tanks, among other things, @GeneralStaffUA reports. \\\\
    \hline
    \\
        \textbf{https://twitter.com/Hajun\_BY/status/1659671266687680513} \\\\
        Report on military activity on the territory of Belarus on May 18 by the Belarusian Hajun: https://motolko.help/en-news/belarusian-hajun-military-activity-on-the-territory-of-belarus-on-may-18/… Online map of military activity in Belarus: http://hajunby.motolko.help 1/12 \\\\
    \hline
    \\
        \textbf{https://twitter.com/NOELreports/status/1636013175005954049} \\\\
        Needs context. These numbers include the promised 100 Leopard 1 (similar to T-72) of which 80-90 can be delivered gradually starting from May this year until autumn. The amount of pledged Leopard 2's are too minimal in numbers to make a difference at the battlefield.\\\\
    \hline
    \end{tabular}
    \caption{Example Tweets (English)}
    \label{tab:example_tweets_english}
\end{table}

\begin{table}[ht]
    \centering
    \begin{tabular}{p{0.95\linewidth}}
    \hline
    \\
        \textbf{https://twitter.com/nos\_osint/status/1511355372002496522} \\\\
        Verwoestingen in Borodyanka, deel 1 geverifieerd: locatie: 50.640540, 29.919766 tijd: video van 4 april \#Borodyanka \selectlanguage{russian} \#Бородянка \selectlanguage{english} \\\\
        \textbf{Google Translate from Dutch:} \\
        Devastation in Borodyanka, part 1 verified: location: 50.640540, 29.919766 time: video from April 4 \#Borodyanka \selectlanguage{russian} \#Бородянка \selectlanguage{english} \\\\
    \hline
    \\
        \textbf{https://twitter.com/YourmediaAgency/status/1497725282249228292} \\\\
        \#ukraine \#russia \#putin UkraineKrieg Tag 3 (26.2.22): Neue Karte (RussMedia). Angriffe der Russen auf Kiew von Norden. Noch kein Vormarsch aufs Zentrum. Schlinge um Mariupol zieht sich immer mehr zu. Putin scheint es nicht um einen schnellen Sieg zu gehen. \\\\
        \textbf{Google Translate from German:} \\
        \#ukraine \#russia \#putin UkraineWar Day 3 (2/26/22): New map (RussMedia). Russian attacks on Kiev from the north. No advance on the center yet. The noose around Mariupol is tightening more and more. Putin doesn't seem to be interested in a quick victory. \\\\
    \hline
    \\
        \textbf{https://twitter.com/radiosvoboda/status/1668833986934898688} \\
        \selectlanguage{ukrainian} Удар РФ по Одесі: троє загиблих, 13 поранених, під завалами можуть бути люди – ОК «Південь» \selectlanguage{english} \\\\
        \textbf{Google Translate from Ukrainian:} \\
        Russian strike on Odesa: three dead, 13 injured, there may be people under the rubble - OK "Pivden" \\\\
    \hline
    \\
        \textbf{https://twitter.com/zloy\_odessit/status/1507696911159078913} \\
        \selectlanguage{russian} Примерно со второй половины марта наблюдается уменьшение интенсивности полётов российских Ан-124 ВКС РФ на аэродром в Мачулищах (Беларусь). В среднем 1-2 рейса в сутки. Зачастую этими рейсами доставлялись ракеты к ОТРК "Искандер". Похоже сказывается дефицит необходимого груза.  \\
        \selectlanguage{english}\textbf{Google Translate from Russian:} \\
        Approximately from the second half of March, there has been a decrease in the intensity of flights of Russian An-124 Russian Aerospace Forces to the airfield in Machulishchi (Belarus). On average 1-2 flights per day. Often these flights delivered missiles to the Iskander OTRK. It looks like there is a shortage of necessary cargo. \\\\
    \hline
    \\
        \textbf{https://twitter.com/Mmzk\_Y/status/1556442432363167744} \\
        \selectlanguage{japanese}\begin{CJK}{UTF8}{min} ロシア人は、ムィコラーイウ地域の弾薬庫で 45,000 トンの弾薬が破壊されたと主張しています。広島が 15 kT TNT 相当の爆発だったことを考えると、誰かがそれに気づいたかもしれないと思います。そして、これはどうやら彼らが主張を修正した後です。 -- ロシア盛り過ぎ。 \end{CJK} \\
        \selectlanguage{english}\textbf{Google Translate from Japanese:} \\
        The Russians claim that 45,000 tons of ammunition were destroyed in an ammunition depot in the Mykolaiv region. Given that Hiroshima was an explosion equivalent to 15 kT TNT, I think someone might have noticed it. And this is apparently after they revised the claim. -- Too much Russia. \\\\
    \hline
    \end{tabular}
    \caption{Example Tweets (Multilingual)}
    \label{tab:example_tweets_multilingual}
\end{table}










  









\end{document}